\documentclass[12pt]{article}

 \usepackage{graphicx,amsfonts,amsmath,amssymb,latexsym}
\usepackage{color}
\newcommand{\nc}{\newcommand}
\def\rr#1{(\ref{#1})}

\nc{\lm}{\lambda} \nc{\ve}{\varepsilon} \nc{\vtheta}{\vartheta}
\nc{\bu}{\mathbf{u}} \nc{\bd}{\mathbf{d}} \nc{\red}{\color{red}}
\nc{\green}{\color{green}} \nc{\blue}{\color{blue}}
\nc{\yel}{\color{yellow}}
 \renewcommand{\title}[1] {%
 \begingroup\begin{center}\vspace{0.0cm}\bf\Large
 \addtolength{\baselineskip}{1mm} #1 \end{center}\endgroup}

 \renewcommand{\author}[1] {%
 \begingroup\begin{center}\vspace{0.2cm}\bf #1 \vspace{0.2cm}
 \end{center}\endgroup}

 \newcommand\pl{{{\otimes}}}
 \newcommand\op{{{\oplus}}}
 \newcommand\bx{\sigma^x}
 \newcommand\by{\sigma^y}
 \newcommand\bz{\sigma^z}
\newcommand\bp {\sigma^+}
\newcommand\bm{\sigma^-}
\newcommand\kp{{\sf{h}}}
\newcommand\tkp{{\tilde{\sf{h}}}}
\newcommand\tbx{{\tilde \sigma}^x}
 \newcommand\tby{{\tilde \sigma}^y}
 \newcommand\tbz{{\tilde \sigma}^z}

\newcommand\tbw{ \tilde{\sf{W}}}
\newcommand\bw{{\sf{W}}}

\newcommand\bq{{\sf{Q}}}
 \newcommand\om{\omega}

 \newcommand\ben{\begin{equation*}}
 \newcommand\ebn{\end{equation*}}
 \newcommand\be{\begin{equation}}
 \newcommand\ee{\end{equation}}
 \newcommand{\ba}{\begin{array}}
 \newcommand{\ea}{\end{array}}

\newcommand\bt{{\sf{T}}}

\newcommand{\eps }{\varepsilon }
\newcommand{\epp}{e^{ip'}}
\newcommand{\eqp}{e^{iq'}}

\newcommand{\ep}{e^{ip}}
\newcommand{\eq}{e^{iq}}

\newcommand{\cp}{   {  P}}
\newcommand{\cpm}{{\cal P}_-}
\newcommand{\cpo}{{\cal P}_0}
\newcommand{\cpl}{{\cal P}_+}
\newcommand{\cq}{{   Q}}
\newcommand{\cqm}{{\cal Q}_-}
\newcommand{\cqo}{{\cal Q}_0}
\newcommand{\cql}{{\cal Q}_+}
\newcommand{\ql}{{ Q}_+}

\newcommand{\qo}{{ Q}_0}

\newcommand{\ppl}{{ P}_+}

\newcommand{\po}{{ P}_0}

\newcommand{\bql}{{ \bar Q}_+}

\newcommand{\bqo}{{\bar Q}_0}

\newcommand{\bpl}{{\bar P}_+}

\newcommand{\bpo}{{\bar P}_0}

\newcommand{\ra}{\rangle}

\begin{document}

\begin{center}{\bf \LARGE Spin operator matrix elements\\ in the   quantum Ising chain: fermion approach}
\end{center}

\bigskip
\begin{center}
{\bf N.~Iorgov$^{1,2}$, V.~Shadura$^{1}$, Yu.~Tykhyy$^{1}$}\\[2mm] {\small
              $^{1}$Bogolyubov Institute for Theoretical Physics, Kiev 03143, Ukraine \\
              $^{2}$Max-Planck-Institut f\"ur Mathematik, Bonn 53111, Germany}

\end{center}

\bigskip

\abstract{
Using some modification of the standard fermion technique we derive factorized formula
for spin operator matrix elements (form-factors) between general eigenstates
of the  Hamiltonian of quantum Ising chain in a transverse field  of finite length.
The derivation is based on the approach recently used to derive factorized formula for
$Z_N$-spin operator matrix elements between ground eigenstates of the  Hamiltonian of
the $Z_N$-symmetric  superintegrable chiral Potts quantum chain.
The obtained factorized formulas for the matrix elements of Ising chain
coincide with the corresponding expressions obtained by the Separation of Variables Method.
}

\bigskip
\noindent
{\bf Keywords:} Quantum Ising chain, form-factors, fermions, Onsager algebra

\bigskip

\section{Introduction}
\label{intro}

During the last decade more attention has been drawn to spin matrix elements in finite Ising systems.
Bugrij and Lisovyy in \cite{BL1,BL2}   has been pointed out that one may write completely factorized
closed expressions for spin matrix elements  and form-factor representation for correlation function
of  the two-dimensional Ising model on finite lattice. The finite volume  form-factors of the  spin field
in the Ising field theory was derived by Fonseca and Zamolodchikov   \cite{Zam}.
The Bugrij--Lisovyy formula for  spin matrix elements on finite lattice
was proved in  \cite{Iorgov1,Iorgov2} by the Separation of Variables Method.
In \cite{Palmer,Hyst} another approach to calculation of spin matrix elements was suggested.
While it allows to get the answer in the continuum limit on the cylinder, the lattice situation appears to be more problematic.

Using the Separation of Variables Method the factorized formulas for spin operator matrix elements
were obtained for the quantum Ising model in a transverse field
\cite{Iorgov1,Iorgov2} and for the XY quantum chain  \cite{Iorgov3}.
The quantum Ising chain (QIC) is the special case ($N=2$) of
the $Z_N$-symmetric  superintegrable chiral Potts quantum chain (SCPC) \cite{GehlenRit}.
In the framework of extended Onsager algebra  \cite{Baxter1}  Baxter   conjectured  the SCPC-spin matrix elements
for the ground state Onsager sectors   \cite{Baxter2}. Independently in the same framework in \cite{gipstZN}
these matrix elements were re-obtained, generalized to arbitrary Onsager sectors and proved.
There it was shown that the matrix elements of spin operators between the eigenvectors of the SCPC
Hamiltonian for arbitrary Onsager sectors can be presented in a factorized form with unknown normalization factors ${\cal N}_{PQ}$
depending only on Onsager sectors but not on particular vectors of these sectors.
The derivation does not use the information on how these Onsager sectors are included in large quantum space of the model.
But in order to calculate scalar factors ${\cal N}_{PQ}$ we need such information. Thus probably the derivation of these factors
will require an information on the Bethe-states of related $\tau_2$-model and $sl(2)$-loop algebra symmetries
\cite{Tar,NDCP,ND,AuYP3,Roan}.

In \cite{Baxter1} Baxter noted that the calculation of the matrix elements of spin operators between eigenvectors of Hamiltonian $H=H_0+\kp H_1$
is greatly simplified if one starts from the calculation of matrix elements of spin operators between eigenvectors of $H_0$
(that is of $H$ at $\kp=0$).
In this paper we use fermion technique to describe the Onsager sectors of QIC.
To identify the bases of Onsager sectors in \cite{gipstZN} with fermion bases we diagonalize $H_0$
by means of dual Jordan--Wigner fermion operators.
As in the case of the duality transformation of 2D Ising model on finite lattice \cite{bushad}, the duality  transformation
for the finite  quantum Ising chain  requires consideration of the periodic
and antiperiodic boundary conditions in a uniform way together with a modification of the standard Jordan--Wigner fermions.
Then derivation of spin matrix elements between eigenstates  of $H_0$ reduces to Cauchy determinant which in turn allows
to extract the factors ${\cal N}_{PQ}=N_{P_0,Q_0}$ in an explicit form.
In order to get matrix elements of spin operator between eigenstates of $H$ we use
Bogoliubov transformation of fermion operators. This transformation of a pair of fermion operators with opposite momenta
is directly related to rotation in a two-dimensional tensor component of Onsager sectors.
Bogoliubov transformation leads to a sum over
the different pairs of fermions which can be summed to get a factorized formula \rr{mat_fin} for matrix elements of spin operator
between eigenvectors of $H$.

This paper is organized as follows: In Sect.~2 we define the Hamiltonian of QIC with unified periodic and antiperiodic boundary conditions,
duality transformation, Jordan--Wigner and dual Jordan--Wigner fermions. In Sect.~3 the derivation
of matrix elements of spin operator between eigenstates of $H_0$ is presented by reducing them to Cauchy determinant. This allows to find
the scalar factors $N_{P_0,Q_0}$ (Appendix~A). In Sect.~4 we define Bogoliubov transformation of fermions and find relation between
eigenstates of $H_0$ and $H_1$. In Sect.~5 we use the results of Sects.~3 and 4 to get factorized matrix elements of spin operator
between eigenstates of $H$. The derivation uses summation formula from Appendix~B. Sect.~6 summarizes our results.

\section{The  quantum Ising chain}

\subsection{The Hamiltonian of the quantum Ising chain}

 The  quantum Ising chain (QIC) of length $L$ with periodic boundary condition
is defined by the Hamiltonian
\be \label{hamP}
H^P = H^P_0+ \kp H^P_1 = -\sum_{k=1}^{L-1} \bx_k \bx_{k+1} -  \bx_L\bx_1 - \kp \sum_{k=1}^L \bz_k\,.
\ee
The same model with antiperiodic boundary condition is governed by the Hamiltonian
\be \label{hamA}
H^A = H^A_0+ \kp H^A_1 = -\sum_{k=1}^{L-1} \bx_k \bx_{k+1} +  \bx_L\bx_1 - \kp \sum_{k=1}^L \bz_k\,.
\ee
Both Hamiltonians include term describing interaction with a transverse magnetic field of strength ${\kp}$.
The space of states of these systems is the $L$-fold tensor product of two-dimensional spaces $V_j$, $j=1,\ldots,L$:
${V} = V_1\pl \cdots \pl V_L$.
  The spin operators
  \[ \sigma^{\alpha}_k\:=\:{\mathbf 1}\;\otimes\ldots\otimes\:{\mathbf 1}\:\otimes
         \underbrace{\:\sigma^{\alpha}\:}_{k-th}\otimes\:{\mathbf 1}\:\otimes\ldots\:\otimes\:{\mathbf 1}, \]
 where $\alpha=x,y,z$,  $k=1, \ldots, L$, and $\sigma^{\alpha}$ are Pauli matrices:
\[\bx = \left(\ba{ll}0\,,& 1\\ 1\,,& 0 \ea\right), \quad
 \by = \left(\ba{ll}0\,,& -i\\ i\,,& \,\,0 \ea\right), \quad
 \bz = \left(\ba{ll}1\,,& \,\,0\\ 0\,,& -1 \ea\right),\]
act non-trivially only  on the space   $V_k$ in the tensor product $V$.
In order to calculate  matrix elements of the spin operator $\sigma_k^x$ in this model it is convenient to  consider  the
 periodic and antiperiodic boundary conditions  in a unified formalism.
 For this aim we introduce additional auxiliary two-dimensional space $V_b$. The space of states becomes
\be\label{vspace}
{\cal V} = V_b\pl V_1\pl \cdots \pl V_L \ee
and unified Hamiltonian is
 \be \label{ham} H = H_0+ \kp H_1,
  \ee
where
\be \label{ham01}   H_0  = -\sum_{k=1}^{L-1} \bx_k \bx_{k+1}
- \bz_b\bx_L\bx_1=-\sum_{k=1}^{L } \bx_k \bx_{k+1}, \quad   H_1=  -   \sum_{k=1}^L \bz_k ,
\ee
with the following boundary conditions for spin operators
 \be\label{bousig}
\sigma^{x}_{L+1}=\sigma^{z}_b\sigma^{x}_{1}, \quad \sigma^{z}_{L+1}= \sigma^{z}_{1},
\ee
  which can be extended to a ``quasiperiodic''  conditions
\be\label{persig}
\sigma^{x}_{k+L}=\sigma^{z}_b\sigma^{x}_{k}, \quad \sigma^{z}_{k+L}= \sigma^{z}_{k}.
\ee
In what follows the main object of our consideration is the Hamiltonian (\ref{ham}) acting on the space \rr{vspace}.
The space of states ${\cal V}$ decomposes into direct sum ${\cal V} = {\cal V}^P\op{\cal V}^A$ of two subspaces
${\cal V}^P$ and ${\cal V}^A$ with eigenvalues $+1$ and $-1$ of $\sigma^{z}_b$, respectively.
Since $\sigma^{z}_b$ commutes with the Hamiltonian $H$ these two subspaces are invariant under the action
of $H$. The restriction of $H$ to these subspaces is $H^P$ and $H^A$, respectively.
The boundary conditions \rr{bousig} become periodic conditions, $\sigma^{x}_{L+1}= \sigma^{x}_{1}$, $\sigma^{z}_{L+1}= \sigma^{z}_{1}$,
or antiperiodic condition, $\sigma^{x}_{L+1}= -\sigma^{x}_{1}$, $\sigma^{z}_{L+1}= \sigma^{z}_{1}$, with respect to the eigenvalue
of $\sigma^{z}_b$.

 The Hamiltonian  $H$  also commutes with the Hermitian operator $\bw$
\be\label{defW}
\bw = - \bz_b \prod_{k=1}^L \bz_k\,.
 \ee
Since $\bw^2=1$,  the eigenvalues of $\bw$ are $w=\pm 1$ and the space $\cal H$ can be decomposed
 into the direct sum ${\cal V} = {\cal V}_{NS}\op{\cal V}_R$ of two subspaces
${\cal V}_{NS}$ and ${\cal V}_R$ corresponding to the eigenvalues $w=-1$ and  $w=+1$, respectively.
The subspace ${\cal V}_{NS}$ (resp. ${\cal V}_R$) is called Neveu--Schwarz sector or  ${NS}$-sector
(resp. Ramond sector or  ${R}$-sector).

With respect to the eigenvalues  of the commuting operators  $\sigma^{z}_b$ and  $\bw $,  the space of states $\cal V$ decomposes
 into the following direct sum:
 \be
{\cal V} = {\cal V}^A_{NS}\op{\cal V}^P_{NS}\op{\cal V}^A_R\op{\cal V}^P_R,
\ee
where, for example, if $|\Psi\ra \in {\cal V}^A_{NS}$ we have
$ \bw |\Psi\ra=-|\Psi\ra$, $\sigma^{z}_b |\Psi\ra=-|\Psi\ra$.
Each of these subspaces have dimension $2^{L-1}$ and is invariant under the action of $H$, $H_0$ and $H_1$ because
they commute with $\sigma^{z}_b$ and  $\bw $.

Let us define the translation operator on $\cal V$
\be\label{tran}
\bt  = \frac 12\left(1+\bz_b+\bz_{1}-\bz_b\bz_{1}\right) \bt_{1,2} \bt_{2,3} \cdots \bt_{L-1,L}\,,
\ee
where the operator
\[
\bt_{k,k+1}=\frac 12\left(1+\bx_k\bx_{k+1}+\by_k\by_{k+1}+\bz_k\bz_{k+1}\right)
\]
swaps the vectors in the spaces $V_k$ and $V_{k+1}$. The translation operator $\bt$
acts on the local spin operators as (see also \rr{bousig})
\be\label{tran_act}
\bt \bz_k = \bz_{k+1} \bt,\quad \bt \bx_k = \bx_{k+1} \bt,\quad
\bt \bz_b = \bz_b \bt,\quad \bt \bx_b = \bx_b \bz_1 \bt
\ee
and satisfies the relations
\[
\bt \bt^+ = 1,\quad \bt^{L}= \frac 12\left(1+\bz_b+W-\bz_b W\right), \quad \bt^{2 L} = 1,\quad \bt \bw = \bw \bt\,.
\]
QIC possesses generalized translation invariance with respect to $\bt$, i.e. $\left[H,\bt\right]=0$.
On the subspace ${\cal V}^P$ the operator $\bt$ turns into standard translation operator.

\subsection{Duality transformation }
The presence of the matrix $\sigma^{z}_b$  in  the Hamiltonian \rr{ham}
results in  some modification of standard duality transformation of QIC.
 We  place the dual Pauli matrices  ${\tilde\sigma}^{\alpha}_{l }$, $l= 1,\ldots$, $L$,
 on the dual chain sites which are located between the neighboring sites of the original chain,  and take the following numbering of dual sites:
 the dual   site  $l$ is located between the sites $l$ and $l+1$ of the original chain.
  We define  the dual Pauli matrices by the following relations:
\[
\tbx_{ l }=\bx_b\prod_{k=1}^{ l }\bz_{{k}}, \quad \tbz_{ l}=
\bx_{l }\bx_{l+1},\quad  \tbx_{1  }=\sigma^x_b \sigma^z_1,\quad\tbx_{L  }=\sigma^x_b\prod_{l=1}^L\sigma^z_l,
\]
\be\label{dusig}
  \tbz_{L  }=\bz_b\sigma^x_1\sigma^x_L,\quad
\tbz_{1  }=\sigma^x_1\sigma^x_2,\quad \tbx_{b  }=\sigma^z_b\sigma^x_1,\quad \tbz_{b  }=\prod_{l=1}^L\sigma^z_l.
\ee
The inverse duality transformation has the form
\[
\bx_{ k }=\tbx_b\prod_{l=k}^{L }\tbz_{{l}}, \quad \bz_{ k}= \tbx_{ k-1 }\tbx_{ k  },\quad  \bx_{L  }
=\tbx_b\tbz_L,\quad\bx_{1  }=\tbx_b\prod_{l=1}^L\tilde\sigma^z_l,
\]
\be\label{odusig}
  \bz_{L }=\tbx_{L-1}\tbx_L,\quad
\bz_{1  }=\tbz_b\tbx_1\tbx_L,\quad \bx_{b  }=\tbz_b\tbx_L,\quad \bz_{b  }=\prod_{l=1}^L\tbz_l.
\ee
Duality transformation interchange $H_0$ and $H_1$ and in terms of the dual Pauli matrices the Hamiltonian $H$ becomes
\be
H=H_0+\kp H_1 =  \kp\left(-\sum_{l=1}^{L }
\tbx_{l }\tbx_{l+1  }-\tkp \sum_{k=1}^{L } \tbz_{{k }}
 \right)= \kp\left({\tilde H}_0 + \tkp {\tilde H}_1\right)= \kp \tilde H,
\ee
where  $\tilde \kp = \kp^{-1}$ and  $\tilde H$ is the dual Hamiltonian
with boundary conditions
\be
    \tbx_{L+1 }  =\tbz_b
\tbx_{1 }, \quad \tbz_{L+1 }= \tbz_{1 }.
\ee

Let us define the dual operator $\tbw$ by
\[
\tbw=-\tbz_b \prod_{l=1}^{L }\tbz_{{l }}\,.
\]
It satisfies the relation
$\tbw= \bw$,
that is  $\bw$ is invariant under the duality transformation.

\subsection{Jordan--Wigner fermion operators }

Usually the first step for diagonalization of QIC Hamiltonian is the introduction of   the Jordan--Wigner (JW) fermion creation and
annihilation  operators. In our case they are
\be\label{ferm}
c_k = \bx_b\cdot\prod_{l=1}^{k-1}\bz_l\cdot \bp_{k},
 \quad
c_k^+ = \bx_b\cdot \prod_{l=1}^{k-1}\bz_l \cdot \bm_{k},
\ee
where $\sigma^{\pm}_{k}= (\sigma^{x}_k \pm  i\sigma^{y}_k)/2$. It is easy to verify that JW  operators (\ref{ferm})  satisfy:

 a) the standard anticommutation relations
\be  \label{comJW}
\{c_k, c_l^+ \}=\delta_{k,l}\,,\quad \{c_k, c_l \}=0\,,\quad \{c_k^+, c_l^+ \}=0\,,
\ee

 b)  the boundary conditions (we take into account \rr{bousig} and $\sigma^{y}_{L+1}=\sigma^{z}_b\sigma^{y}_{1}$)
 \be\label{bouJW}
c_{L+1}= \bw c_{1}, \quad c^+_{L+1}=\bw c^+_{1}\,,
\ee

 c) the ``quasiperiodic''  conditions
\be\label{quperJW}
c_{k+L}=\bw c_k, \quad c^+_{k+L}=\bw c^+_k\,,
\ee

 d) action of  the translation operator $\bt$ consistent with \rr{quperJW}
\be\label{tran_ferm}
\bt c_{k} = c_{k+1}\bt,\quad \bt c_{k}^+ = c_{k+1}^+ \bt\,.\ee

In terms of the fermion number operator $Q =\sum_{k=1}^{L}c^{+}_{k}c_{k}$ we have the relations
 \be \label{bwfer}
\bw=- \bz_b \prod_{k=1}^L \bz_k=- \bz_b \prod_{k=1}^{L}\exp(i\pi c^{+}_{k}c_{k})=- \bz_b (-1)^{ Q },
 \ee
\be \label{combwJW}
\left[\bw, c_k\right] =0\,, \quad \left[\bw, c^+_k\right] =0\,.
 \ee

The operators $H_0$ and  $H_1$ from (\ref{ham01})  in terms of
the JW fermion operators take the form
\be\label{ham0c}
H_0= -\sum_{k=1}^{L } \bx_k \bx_{k+1}  =-\sum_{k=1}^{L }(c^{+}_{k }-c_{k })(c^{+}_{k+1 }+c_{k+1}),
\ee
\be\label{ham1c}
 H_1 = -   \sum_{k=1}^L \bz_k=
       2\sum_{k=1}^{L}\left(c^{+}_{k}c_{k}-\frac 12\right)=
    \sum_{k=1}^{L}  (c^{+}_{k}-c_{k})(c^{+}_{k}+c_{k}).
      \ee
The last equation gives that JW fermions diagonalize $H_1$.

\subsection{ Dual Jordan--Wigner fermion operators}

As discussed in the Introduction in order to calculate the spin operator matrix elements in  the QIC
we must first solve the auxiliary problem, namely, to calculate the matrix element in the basis of eigenvectors of the Hamiltonian $H_0$,
which is $H$ at $\kp=0$.
In order to diagonalize $H_0$ it is convenient to define the dual JW fermion creation  and annihilation operators $a^+_{k}$, $a_{k}$,
$k=1,\ldots,L$, through the relations
\be \label{induJW}
c^+_k-c_k =-(a^+_{k }-a_{k }),\quad c^+_{k+1}+c_{k+1}=a^+_{k }+a_{k }
\ee
interchanging $H_0$ and $H_1$ given by (\ref{ham0c}) and  (\ref{ham1c}):
\be\label{dham0}
H_0= -\sum_{k=1}^{L }(c^{+}_{k }-c_{k })(c^{+}_{k+1 }+c_{k+1 })=
\sum_{k=1}^{L}(a^+_{k}-a_{k})(a^+_{k}+a_{k})=2 \sum_{k=1}^L\left(a_{k }^+ a_{k } -  \frac 12\right)\,,
\ee
\be\label{dham1}
H_1= \sum_{k=1}^{L}(c^{+}_{k}-c_{k})(c^{+}_{k}+c_{k})=
  -\sum_{k=1}^{L}(a^{+}_{k+1 }-a_{k+1 })(a^{+}_{k }+a_{k })\,.
\ee

{}From \rr{induJW}, we have the explicit formulas for the dual JW fermion operators
\be\label{df}
a_{k}=\frac 12 \left(c^+_{k+1}+c_{k+1}+(c^+_k-c_k) \right)\,,\quad
a^+_{k}=\frac 12 \left(c^+_{k+1}+c_{k+1}-(c^+_k-c_k) \right)\,.
\ee
Also these operators can be written in terms of $\sigma^{\alpha}_k$
\be
\label{duJWsig}
a_{k} =\frac 12\bx_b \cdot \prod_{l=1}^k\bz_l\cdot \left(\sigma^x_{k+1}-\sigma^x_{k}\right)\,,\quad
a^+_{k}=\frac 12\bx_b \cdot \prod_{l=1}^k\bz_l\cdot \left(\sigma^x_{k+1}+\sigma^x_{k}\right)\,.
 \ee
or in terms of the dual operators ${\tilde\sigma}^{\alpha}_k$
\be \label{duduJWsig}
a_{k  }= \tbx_b \cdot \prod_{l=k+1}^{L}\tbz_{l } \cdot \tilde\sigma^+_k\,, \quad
a^+_{k }= \tbx_b \cdot \prod_{l=k+1}^{L}\tbz_{l } \cdot  \tilde\sigma^-_k\,,
\ee
where $\tilde\sigma^\pm_k=( \tbx_{k }\pm {i}\tby_{k })/2$ and $\tby_{k }=i \tbx_{k }\tbz_{k }$\,.

Due to the linearity of the transformation \rr{df} and using the properties
(\ref{comJW})--(\ref{combwJW}) of the JW operators
$c^+_l$, $c_l$ together with (\ref{duJWsig}), a straightforward calculation gives for the dual JW operators:

  a) the   anticommutation relations
  \be\label{comduJW}
\{a_k, a_l^+ \}=\delta_{k,l}\,,\quad \{a_k, a_l \}=0\,,\quad \{a_k^+, a_l^+ \}=0\,,
\ee

 b)  the boundary conditions
 \be\label{bouduJW}
a_{L+1}= \bw a_{1}\,, \quad a^+_{L+1}=\bw a^+_{1}\,,
\ee

 c)      the ``quasiperiodic''  conditions
\be\label{quperduJW}
a_{k+L}=\bw a_k\,, \quad a^+_{k+L}=\bw a^+_k\,,
\ee

d) the commutativity with the operator  $\bw=\tbw$
\be \label{bwduJW}
\left[\bw,a_{k }\right] = \left[\bw, a^+_{k }\right] =0\,,
 \ee

e) the anticommutativity with the operator $\bz_b$
\be \label{comdusigJW}
\{\bz_b, a_{k}\} =0, \quad\{\bz_b, a^{+}_{k}\} =0\,,
\ee

f) action of  the translation operator $\bt$ consistent with \rr{quperduJW}
\be\label{tran_duferm}
\bt a_{k} = a_{k+1}\bt,\quad \bt a_{k}^+ = a_{k+1}^+ \bt\,.\ee

{}From \rr{bwduJW} it follows that the subspaces ${\cal V}_{NS}$ and ${\cal V}_R$ are invariant with respect
to the action of fermion algebra \rr{comduJW} having unique irreducible representation --- standard Fock representation
of dimension $2^L$. Since the dimensions of ${\cal V}_{NS}$ and ${\cal V}_R$ are also $2^L$, the action of the algebra \rr{comduJW}
on each of them is irreducible. We will denote the corresponding vacuum states in the irreducible subspaces ${\cal V}_{NS}$ and $ {\cal V}_{R}$
by $|{0}\ra_{0,NS}$ and $|{0}\ra_{0,R}$, respectively. Using \rr{defW} and \rr{duJWsig} it is easy to verify  that the following two
vectors
\be\label{explNSvac}
| 0\rangle_{0,NS} = \frac{1}{{2}^{(L+1)/2}}\left[\left(\ba{l}1\\0\ea\right)^{(b)}
\left(\ba{l}1\\1\ea\right)^{(1)}
...\left(\ba{l}1\\1\ea\right)^{(L)}
 +
 \left(\ba{l}1\\0\ea\right)^{(b)}\left(\ba{l}1\\-1\ea\right)^{(1)}
...\left(\ba{l}1\\-1\ea\right)^{(L)}\right],
\ee
\be\label{explRvac}
|0\rangle_{0,R} = \frac{1}{{2}^{(L+1)/2}}\left[\left(\ba{l}1\\0\ea\right)^{(b)}\left(\ba{l}1\\1\ea\right)^{(1)}
...\left(\ba{l}1\\1\ea\right)^{(L)}
 -   \left(\ba{l}1\\0\ea\right)^{(b)}\left(\ba{l}1\\-1\ea\right)^{(1)}
...\left(\ba{l}1\\-1\ea\right)^{(L)}\right]
\ee
belonging to ${\cal V} = {\cal V}_{NS}\oplus {\cal V}_{R}=V_b\pl V_1\pl \cdots \pl V_L$ satisfy the necessary equations for these vacuum states:
\[ \bw | {0}\ra_{0,NS}=-| {0}\ra_{0,NS},  \qquad a_{l }| {0}\ra_{0,NS}=0\,,\quad l=1,\ldots,L, \]
\[ \bw | {0}\ra_{0,R}=| {0}\ra_{0,R},\qquad a_{l }| {0}\ra_{0,R}=0\,,\quad l=1,\ldots,L\,.\]
Also we have $\bt | 0\rangle_{0,NS} =| 0\rangle_{0,NS}$ and $\bt | 0\rangle_{0,R} =| 0\rangle_{0,R}$

Because of \rr{bwduJW} we may restrict operators $a_l$, $a_l^+$, $l=1,\ldots,L$, to the Neveu--Schwarz sector ${\cal V}_{NS}$
(resp. to the Ramond  sector $ {\cal V}_{R}$)
where $\bw$ has eigenvalue  $-1$ (resp. $+1$). The restricted operators will be denoted $f_l$, $f^{+}_l$
(resp. $d_l$, $d^{+}_l$) which have due to \rr{bouduJW} antiperiodic (resp. periodic) boundary condition:
\be\label{bcfd}
f^{+}_{L+l}=-f^{+}_l\,,\qquad d^{+}_{L+l}=d^{+}_l \,.
\ee
The set of the vectors
\be\label{efs}
\left(f_1^{+}\right)^{n_1} \left(f_2^{+}\right)^{n_2} \cdots
\left(f_L^{+}\right)^{n_L} |0\ra_{0,NS}, \quad
\left(d_1^{+}\right)^{n_1} \left(d_2^{+}\right)^{n_2} \cdots
\left(d_L^{+}\right)^{n_L}|0\ra_{0,R}\,,\quad n_k\in \{0,1\}\,,
\ee
constitute a basis of ${\cal V} = {\cal V}_{NS}\oplus {\cal V}_{R}$
on which the operator \rr{dham0} is diagonal.

As it was explained in the Sect.~2.1, the boundary condition of $H$ and $H_0$ is fixed by the eigenvalues of the operator $\bz_b$.
Using  \rr{duduJWsig} one can relate the dual fermion number operator $\tilde\bq$ to $\bz_b$:
\be\label{bz_tildeQ}
 \bz_b=\prod_{l=1}^L\tbz_{l } =  (-1)^{\tilde\bq }\,,\qquad
\tilde\bq=\sum_{k=1}^{L}a^{+}_{k}a_{k}\,.
\ee
Hence the states from \rr{efs} with an even number of excitations (including the vacuum states $|{0}\ra_{0,NS}$ and $|{0}\ra_{0,R}$)
belong to ${\cal V}^P$ and  the states with odd number of excitations belong to ${\cal V}^A$.

Let us define the momentum representation for the dual Jordan--Wigner fermion operators.
It is defined by means of the discrete Fourier transformation:
\be\label{ftf}
{f}_q = \frac{1}{\sqrt{L}}\sum_{l=1}^L {f}_l e^{-i  q \,l}, \quad
{f}_q^+ = \frac{1}{\sqrt{L}}\sum_{l=1}^L {f}_l^+ e^{i q \,l},\ee
\be\label{ftd}
{d}_p = \frac{1}{\sqrt{L}}\sum_{l=1}^L {d}_l  e^{-i p \,l}, \quad
{d}_p^+ = \frac{1}{\sqrt{L}}\sum_{l=1}^L {d}_l^+ e^{i p \,l},\ee
where in the $NS$-sector, due to the antiperiodicity \rr{bcfd},
the momentum  $q$ takes ``half-integral''    values $q=2\pi(k+1/2)/L$, $k\in \mathbb{Z}$,
and in the $R$-sector, due to the periodicity \rr{bcfd},
the momentum $p$ takes ``integral'' values $p =2\pi k/L$, $k\in \mathbb{Z}$.
All the momenta are defined up to a multiple of $2\pi$. The set of ``half-integral'' (resp. ``integral'') momenta $q$ with $-\pi<q\le \pi$
will be denoted $\cal Q$ (resp. ${\cal P}$). All the momenta enter $\cal Q$ (resp. ${\cal P}$) by pairs (i.e. two different momenta
$q,-q\in {\cal Q}$) except the momenta $0$ and $\pi$.
The momentum $0\in\cal P$, the momentum $\pi\in {\cal P}$ if the length $L$ of the chain is even and  $\pi\in {\cal Q}$ if $L$ is odd.
In what follows we will use the basis
\be\label{esNS}
| {   Q }\rangle_{0,NS}= |{q_1},\ldots, \,{q_m}\ra_{0,NS} =
{f}^+_{q_1}  {f}^+_{q_2}\cdots{f}^+_{q_m}|0\ra_{0,NS}\in {\cal V}_{NS}\,,
\ee\be\label{esR}
| {   P }\rangle_{0,R}= |{p_1},\ldots, \,{p_n}\ra_{0,R} =
{d}^+_{p_1}  {d}^+_{p_2}\cdots{d}^+_{p_n}|0\ra_{0,R}\in {\cal V}_{R}
\ee
of $\cal V$ instead of basis \rr{efs}. These states are labeled by the subsets
$Q\subset{\cal Q}$ and $P\subset{\cal P}$ of excited fermion momenta in
NS-sector and R-sector, respectively.
We introduce the subsets $Q_+$, $Q_-$, $Q_0\subset Q$ as
\[
Q_+=\{q\in Q\,|0<q<\pi\,,\,  -q\in Q  \}, \  Q_-=\{q\in Q\,|-\pi<q<0\,,\,  -q\in Q \}, \  Q_0=
Q\backslash (Q_+\cup Q_-)
\]
and the subsets $\bar Q_+$, $\bar Q_-$, $\bar Q_0\subset \bar Q={\cal Q}\backslash Q$ by similar formulas with the replacement
$Q\to \bar Q$. The cardinality of the sets $Q$, $Q_+$, etc. are $m$, $m_+$, etc., respectively.
We also will use the sets $\cql=\ql\cup\bql$,  $\cqo=\qo\cup \bqo$, $\cqm=Q_-\cup\bar Q_-$.
Analogously we introduce the subsets $P$, $P_+$, etc. of ${\cal P}$ which have the cardinality $n$, $n_+$, etc.
and the sets $\cpl=P_+\cup\bar P_+$,  $\cpo=\po\cup \bpo$, $\cpm=P_-\cup\bar P_-$.
We have the relations
\[
L= m+\bar m=n+\bar n, \quad m=2 m_+  + m_0,\quad  \bar m= 2\bar m_+  +  \bar m_0,\quad
n= 2n_+    + n_0,\quad \bar n= 2\bar n_+  +  \bar n_0.
\]

Formulas \rr{ftf} and \rr{ftd} imply that the translation operator $\bt$ acts diagonally on fermion operators in momentum representation:
\be\label{tran_duferm_mom}
\bt f_{q} = e^{i q} f_{q} \bt,\quad \bt f_{q}^+ = e^{-i q} f_{q}^+ \bt\,,\quad
\bt d_{p} = e^{i p} d_{p} \bt,\quad \bt d_{p}^+ = e^{-i p} d_{p}^+ \bt\,.\ee
Using \rr{dham0} and \rr{dham1} we can rewrite restrictions of $H_0$ and $H_1$ on NS and R-sectors
in terms of dual Jordan--Wigner fermion operators
in momentum representation:
\be\label{H01mom}
H_0 = \sum_q(2{a}_q^+ {a}_q-1)\,,\qquad H_1= \sum_q\left((1-2 {a}_q^+ {a}_q )
\cos  q+i\left({a}_q^+ {a}_{-q}^+ +{a}_q {a}_{-q} \right)\sin q\right),
\ee
where for $NS$-sector ${a}_q={f}_q$, $q\in {\cal Q}$,  and for  $R$-sector
${a}_q={d}_q$,  $q\in {\cal P}$.
In what follows we will often use the same notations $H$, $H_0$ and $H_1$ for the restrictions of the corresponding Hamiltonians
\rr{ham}, \rr{ham01} to subspaces ${\cal V}_{NS}$ and ${\cal V}_R$.

\section{Spin operator matrix elements \\
between the eigenstates of the Hamiltonian $ H_0$}

In this section we will find the matrix elements of spin operator $\bx_1$ between
eigenstates \rr{esNS} and \rr{esR} of $H_0$.
The relation $\bw \bx_1=-\bx_1\bw$ gives that the action of $\bx_1$ swaps two eigenvalues $\pm 1$ of $\bw$,
that is  the operator $\bx_1$ maps ${\cal V}_{NS}$ to ${\cal V}_{R}$ and vice versa.
The operator $\bx_1$ commutes with $H_0$ from \rr{ham01} and therefore due to \rr{dham0} it commutes with
the dual fermion number operator $\tilde\bq$ from \rr{bz_tildeQ}. Hence action of $\bx_1$ conserve
the number of excitations. In particular, it should map $|0\ra_{0,NS}$  to a vector proportional to $|{0}\ra_{0,R}$.
Using explicit formulas \rr{explNSvac} and \rr{explRvac} we get
\be\label{vac0ME}
| {0}\ra_{0,R}=\bx_1|  0\ra_{0,NS}\,,\quad |  0\ra_{0,NS}=\bx_1 | {0}\ra_{0,R}\,.
\ee
Using \rr{duJWsig} we get the commutation relations
\[
a_{l}\bx_1+\bx_1a_{l}=  0, \quad a^+_{l}\bx_1+\bx_1 a^+_{l} =0\,, \quad l=1,2,\ldots,L,
\]
\be\label{trdufe}
  \bx_1 {d}_k \bx_1=  -{f}_k, \quad
\bx_1{d}^{+}_l \bx_1  =  -{f}^{+}_l \,\quad l=1,2,\ldots,L,\ee
which give after Fourier transformations \rr{ftf} and \rr{ftd}
\be\label{dsigx}
 \bx_1 {d}_p \bx_1 = -\frac{ 1}{\sqrt{L}}\sum_l {f}_l e^{-i p\,l}
 = -\frac{ 1}{L}\sum_k\sum_q {f}_q e^{i(q-p)\, k}=  \frac{2}{L}\sum_q {f}_q
\frac{1}{1- e^{i(p-q)}}\,,\ee
\be\bx_1 {d}_p^+ \bx_1 = -\frac{ 1}{\sqrt{L}}\sum_l {f}_l^+ e^{i p\,l}
 = -\frac{ 1}{L}\sum_k\sum_q {f}_q^+ e^{-i(q-p)\, k}=  \frac{2}{L}\sum_q {f}_q^+
\frac{1}{1- e^{-i(p-q)}}\,,
\ee
allowing to calculate matrix elements ${}_{0,R}\langle{P}| \bx_1|{Q}\rangle_{0,NS}$ of the spin operator   $\bx_1$
between vectors \rr{esNS} and \rr{esR}. Since $\bx_1$ conserve the number of excitation, the non-zero matrix elements will appear only if $n=m$:
\[{}_{0,R}\langle{{p_1},\ldots, \,{p_n}}| \bx_1|{{q_1},\ldots, \,{q_m}}\rangle_{0,NS}=
{}_{0,R}\langle 0 |{d}_{p_n} \cdots {d}_{p_2} {d}_{p_1}\bx_1 {f}^+_{q_1}
 {f}^+_{q_2} \cdots {f}^+_{q_m}|0\ra_{0,NS}=
\]
\[ = \frac{2}{L}\sum_{q\in {\cal Q}}
 \frac{1}{1-e^{-i(q-p_1)}} \,{}_{0,R}\langle 0 |
 {d}_{p_{n }} {d}_{p_{m-1}} \cdots  {d}_{p_2}  \bx_1 f_q f^+_{q_1} {f}^+_{q_2} \cdots f^+_{q_m}|0\ra_{0,NS} =
\]
\be = \delta_{n,m}\frac{2}{L}\sum_{k=1}^m
 \frac{(-1)^{k-1}}{1-e^{-i(q_k-p_1)}}{}_{0,R}\,\langle
0 | {d}_{p_{m }} {d}_{p_{m-1}}\cdots  {d}_{p_2}  \bx_1
f^+_{q_1} f^+_{q_2} \cdots f^+_{q_{k-1}} f^+_{q_{k+1}} \cdots
f^+_{q_m}|0\ra_{0,NS}.
\ee
Taking into account ${}_{0,R}\langle 0|\bx_1|0\ra_{0,NS} = 1$ which follows from \rr{vac0ME}, we
see that we get recurrent relation for the expansion of the determinant
\be\label{det1}
{}_{0,R}\langle 0 |{d}_{p_m} \cdots  {d}_{p_2} {d}_{p_1}\bx_1
 {f}^+_{q_1}
 {f}^+_{q_2} \cdots {f}^+_{q_m}|0\ra_{0,NS}
=\delta_{n,m}\,
\det
\left(\ba{llll}A_{1,1},&A_{1,{2}},&...,&A_{1,m}\\
A_{2,1},&A_{2,{2}},&...,&A_{2,m}\\
...\,,&...\,,&...\,,&...\\
A_{m,1},&A_{m,{2}},&...,& A_{m,m}\ea\right),\ee
with the matrix elements
\be A_{i,j} =
\left(\frac{2}{L}\right)\frac{1}{1-e^{i(p_i-q_j)}}= \left(\frac{2}{L}\right)\frac{e^{iq_j}}{e^{iq_j}-e^{i p_i }}\,,\ee
with respect to first row. It is a variant of Wick theorem with two-particle pairing given by $A_{i,j}$.
The determinant \rr{det1} can be transformed to the Cauchy determinant for which we have
\[
\det\left( \frac{1}{x_i-y_j}\right)_{1\leq i,j\leq n}=\frac{\prod_{1\leq i<j\leq n}\left(x_j-x_i\right)
\left(y_i-y_j\right)}{\prod_{i,j=1}^{n}\left(x_i-y_j\right)}.
\]
Introducing the variables $x_j =e^{i\,q_j}$, $y_i =  e^{i\,p_i}$ we get
\[
 {}_{0,R}\langle{  P }| \bx_1| {   Q }\rangle_{0,NS}= {}_{0,R}\langle {p_1}, \,{p_2}, \ldots , {p_m}|\bx_1 |{q_1},\,
 {q_2}, \ldots, \,{q_m}\ra_{0,NS} =
\]
 \be \label{mat_IW_fer}
= \delta_{n,m}(-1)^{m(m-1)/2}
 \left(\frac{2}{L}\right)^{m}
 \frac{\prod_{i<j}^m\left(e^{i\,q_i}-e^{i\,q_j}\right)\prod_{i<j}^m\left(e^{i\,p_i}-e^{i\,p_j} \right)
 \prod_{j=1}^m e^{i\,q_j}} {\prod_{i,j}^m \left(e^{i\,q_i}-e^{i\,p_j} \right)}.
\ee

For convenience, we will use the following ordering of the momenta in the state
  $|\cq\ra_{0,NS}=|{q_1},\, {q_2}, \ldots, \,{q_m}\ra_{0,NS}$:
  from the beginning the pairs $(-q, q)$, for $q\in Q_+$, are going on and then $q\in Q_0$ are going on
in an order fixed over all this paper. The final formula for the matrix elements \rr{mat_fin} will not depend on this ordering.
  Such ordering defines the sign of $|\cq\ra_{0,NS}$ uniquely.
Analogously we define the ordering of the momenta for $|P\rangle_{0,R}$.
It is obvious that the sets  $\ql$, $\qo$  and  $\ppl$, $\po$
  completely characterize  the states    $|Q\rangle_{0,NS}$  and  $|P\rangle_{0,R}$.
Hence we can use the following notations:
\be\label{stateH0}
 | { \ql, \qo }\rangle_{0,NS}=| {   Q }\rangle_{0,NS}\,,
 \quad | { \ppl, \po }\rangle_{0,R}=| { P  }\rangle_{0,R}\,.
\ee
The operators $H_0$ and $H_1$ generate Onsager algebra. From \rr{H01mom} it follows that the space ${\cal V}_{Q_0}\subset {\cal V}_{NS}$
of all the states with the same fixed $Q_0$ is invariant with respect to the action of $H_0$ and $H_1$  and in fact it is space of an irreducible representation of this Onsager algebra.
The basis vectors in ${\cal V}_{Q_0}$ are labeled by the sets $Q_+$ and the dimension of ${\cal V}_{Q_0}$ is $2^{|{\cal Q}_+|}$.
The space ${\cal V}_{Q_0}$ can be identified with the tensor product of $|{\cal Q}_+|$ two-dimensional spaces
with each tensor component labeled by $q\in {\cal Q}_+$.
To vector $i^{-m_+}|\ql, Q_0\rangle_{0,NS}$ we put in correspondence the tensor product of
two-dimensional vectors in each tensor component: $(1,0)^T$ if $q\in \bar Q_+$ and  $(0,1)^T$ if $q\in Q_+$.
Then from \rr{H01mom} it follows that
in the basis of vectors $i^{-m_+}|\ql, Q_0\rangle_{0,NS}$ with above mentioned identification of bases
the Hamiltonian $H$ has the form of Eq.~(10) from \cite{gipstZN}.
We have analogous situation  for Onsager sector ${\cal V}_{P_0}\subset {\cal V}_{R}$.

The results of \cite{gipstZN} predict the matrix elements of spin operator up to unknown constants $N_{\po,\qo}$
depending only on Onsager sectors:
\[
 {}_{0,R}\langle {\cp}|\bx_1 |{\cq}\ra_{0,NS}=
 {}_{0,R}\langle  \ppl , \po| \bx_1| {\ql, \qo }\rangle_{0,NS}=
i^{(n_0-m_0)/2}\delta_{2|I|,L-m_0-{\bar n}_0 }N_{\po,\qo}\times
\]
\be\label{mat_fin_0}\times
\frac{(-1)^{{  n}_+}
\prod_{\alpha\in {\bar I}}\tan{\frac{\alpha}{2}}}
{\prod_{\alpha\in I}(\cos\alpha+1)^\sigma\prod_{\beta\in
{\bar I}}(\cos\beta-1)^\tau\prod_{\alpha\in I}\prod_{\beta\in
{\bar I}}(\cos\alpha-\cos\beta)},
\ee
where
\[I =\ql\cup\bpl  ,\quad {\bar I} = \ppl\cup\bql,\quad |I|=m_++{\bar n}_+\,,
\]\[
L=2m_++2\bar m_++m_0+\bar m_0, \quad 2|I|-L+\bar n_0+  m_0=m-n,
\]
\be\label{defsigtau}
2\tau= -P_{0,0}+{\bar P}_{0,0} + Q_{0,0} - {\bar Q}_{0,0}+1,\quad
2\sigma=-Q_{0,\pi}+{\bar Q}_{0,\pi}+P_{0,\pi}-{\bar P}_{0,\pi}+1,
\ee\be\label{defP0Ppi}
P_{0,0}{=} \left\{\ba{l}1,\ 0\in P_0\\0,\ 0\not\in P_0
\ea,\right.\, P_{0,\pi}{=} \left\{\ba{l}1,\ \pi\in P_0\\0,\ \pi\not\in P_0 \ea,\right.\,
\bar P_{0,0} {=}\left\{\ba{l}1, 0\in \bar P_0\\0, \  0\not\in \bar P_0 \ea,\right.\,
\bar P_{0,\pi} {=} \left\{\ba{l}1, \ \pi\in \bar P_0\\0, \ \pi\not\in \bar P_0\ea\,.\right.
\ee
The functions $Q_{0,0}$, $Q_{0,\pi}$, $\bar Q_{0,0}$,   $\bar Q_{0,\pi}$ are defined similarly.
Note that always we have $Q_{0,0}=\bar Q_{0,0}=0$ but we included them for uniformity.
{}From \rr{defsigtau} it follows $\sigma,\tau\in\{0,1\}$.
Also we used  $\delta_{m,n}=\delta_{m_0-n_0,2(n_+-m_+)}=\delta_{2|I|,L-m_0-{\bar n}_0}$.

Formula \rr{mat_IW_fer} fixes $N_{\po,\qo}$ in \rr{mat_fin_0} uniquely.
The derivation in Appendix A gives independent proof of the formula \rr{mat_fin_0}
(it is the identity  (\ref{idenA1}) at $n=m$) and allows to fix $N_{\po,\qo}$ in different equivalent ways.
One possible expression for $N_{\po,\qo}$ is given by
\[
N_{\po,\qo}=\delta^{(\text{mod}\, 2)}_{m_0-n_0,0}
\left(\frac{2}{L}\right)^{(m_0+n_0)/2}N^0_-\,N_2\, N_e\, A_{\po,\qo}\times
\]
\be \label{npqA}
\times\prod_{q\in \cql}\frac{1}{\tan(q/2)}
\prod_{q\in \cql}(\cos q-1)^{\tau}
\prod_{p\in \cpl}(\cos p+1)^{\sigma}\prod_{p\in \cpl}\prod_{q\in \cql}  (\cos p-\cos q)\,,
\ee
where
\[
N^0_-=(-1)^{n_0(n_0-1)/2} (-1)^{(m_0-n_0)(m_0-n_0-2)/8} (-1)^{(L-m_0-\bar n_0)(n_0-m_0)/4},
\]
\[
N_2= 2^{-(n_0-m_0)^2/4},\quad
N_e=\prod_{p\in \po}e^{ip(m_0-n_0)/2}\prod_{q\in \qo}e^{-iq(m_0-n_0-2)/2},
\]\[
A_{\po,\qo}=\frac{\prod_{p<p'\in \po} (\ep-\epp)
\prod_{q<q'\in \qo} (\eq-\eqp)}{
\prod_{p\in \po}\prod_{q\in \qo}
( \eq-\ep)}\,,\quad
\delta^{\,(\text{mod}\, 2)}_{m_0-n_0, 0  } = \frac 12\left((-1)^{n_0-m_0}+1\right).
\]
It can be obtained from \rr{idenA1} by fixing $P_+$ and $Q_+$ as empty sets.
This expression is explicitly does not depend on particular vectors of Onsager sectors.
This expression for $N_{\po,\qo}$ will not be used for the calculations in the paper.
Another more useful equivalent expression for $N_{\po,\qo}$ is given by \rr{npqH}.
Now we can substitute the obtained $N_{\po,\qo}$ to the formulas of Sect.~4.2 and Sect.~4.3 of
\cite{gipstZN} (they correspond to formula \rr{f_after_sum} of present paper) and obtain  after some regrouping of factors
 the factorized matrix elements of spin operator between eigenvectors of $H$.
Instead of this direct way we choose more long but self-contained presentation of results
and show also that the rotations in representations of Onsager algebra correspond to Bogoliubov transformations
of pairs of fermions with opposite momenta.

\section{ Eigenvectors of the quantum Ising chain}

Using \rr{H01mom} we can rewrite the initial Hamiltonian $H = H_0+ \kp H_1$ in terms of dual Jordan--Wigner fermion operators
in the momentum representation:
\be \label{hamIq}
   H = \sum_q\left((2{a}_q^+ {a}_q-1)(1-\kp\cos q)+i\left({a}_q^+ {a}_{-q}^+ +{a}_q {a}_{-q}
\right)\kp\sin q\right),
\ee
where for $NS$-sector ${a}_q={f}_q$, $q\in {\cal Q}$,  and for  $R$-sector
${a}_q={d}_q$,  $q\in {\cal P}$.
Bogoliubov transformation defines new fermion operators, $0\le q\le \pi$:
\[b_q = c_{q} a_q + i s_{q} a_{-q}^+, \quad b_q^+ = c_{q} a_q^+ - i s_{q} a_{-q},\]
\[b_{-q} = c_{q} a_{-q} - i s_{q} a_{q}^+, \quad b_{-q}^+ = c_{q} a_{-q}^+ + i s_{q} a_{q},\]
 which diagonalize the Hamiltonian (\ref{hamIq})
\be \label{hamIb}
 H  = 2 \sum_q\varepsilon_q\left(  {b}_q^+ {b}_q  -\frac 12\right),
\ee
where in the ferromagnetic regime, $0\le \kp<1$, we have
\be\label{eeq}
  \varepsilon_q=\sqrt{1+\kp^2-2\kp\cos q}\,,\quad  \ve_0=1-\kp\,,\quad \ve_\pi=1+\kp\,,
\ee
\be\label{cq}
c_{q}=\cos \frac{\theta_q}{2},\quad s_{q}=\left|\sin\frac{\theta_q}{2}\right|,
\quad \tan\theta_q=\frac{\kp\sin q}{1-\kp\cos q}.
\ee

Note that   $c_{q}$ and $s_{q}$ can be represented in the form
$$c_q(\varepsilon_q,\kp) =  \sqrt{\frac{(\eps_\pi+\varepsilon_q )(\eps_0+\varepsilon_q )}{4 \varepsilon_q}},\quad
s_{q}(\varepsilon_q,\kp) = \frac{1}{2\varepsilon_q}\sqrt { {(\eps_\pi-\varepsilon_q )
( \varepsilon_q-\eps_0)}{  \varepsilon_q}}.$$
Formally we have
\be\label{meps}
c_q(-\varepsilon_q,\kp) =s_{q}(\varepsilon_q,\kp),\quad s_{q}(-\varepsilon_q,\kp)=
-c_q(\varepsilon_q,\kp)\,.
\ee

The vacuum states $|0\rangle_{NS}$ and $|0\rangle_{R}$ for the Hamiltonian $H$ in $NS$- and $R$-sector
\be\label{vacI}
  |0\rangle_{NS}  = \prod_{0<q <\pi }\left(c_q + i s_q {f}_{-q}^+ {f}_q^+\right)
|0\rangle_{0,NS}\,,\quad
 |0\rangle_{R}  = \prod_{0<q <\pi } \left(c_q + i s_q {d}_{-q}^+ {d}_q^+\right)
|0\rangle_{0,R}\,,
\ee
where $|0\rangle_{0,NS}$ and $|0\rangle_{0,R}$ are the vacuum states  of the Hamiltonian  $H_0$
in corresponding sectors.
 It is easy to show that for both sectors the following relations are fulfilled: ${b}_q|0\rangle=0$ and
   \be \label{bact}
b^+_q\left(c_q + i s_q {a}_{-q}^+ {a}_q^+\right)|0\rangle_0=a^+_q|0\rangle_0,\quad
b^+_{-q}b^+_{q}\left(c_q + i s_q {a}_{-q}^+ {a}_q^+\right)|0\rangle_0= (c_q a_{-q}^+ a_{q}^+ + i s_q)|0\rangle_0. \ee

As in the case of the eigenvectors of $H_0$ we will label the eigenvectors of $H$ by the same type of sets of momenta
and use the same ordering. For example, in the NS-sector the vectors are labeled by a set $Q$ of momenta:
\[|{Q}\rangle_{NS}= |{q_1},\ldots, \,{q_m}\ra_{NS} =
{b}^+_{q_1}  {b}^+_{q_2}\cdots{b}^+_{q_m}|0\ra_{NS}\in {\cal V}_{NS}\
\]
or by two sets $Q_+$ and $Q_0$ with the following ordering of the momenta in the state:
\[ |{q_1},\,
 {q_2}, \ldots, \,{q_m}\ra_{NS}  =| \cq\rangle_{NS}  = |Q_+,Q_0\rangle_{NS}=
 \prod_{q\in Q_+} \left(b^+_{-q} b_q^+\right) \prod^{\longrightarrow}_{q \in Q_0}b_q^+   |0\rangle_{NS}\,.
\]
The formulas \rr{tran_duferm_mom} for the action of translation operator
together with the formulas for Bogoliubov transformation imply
\[\bt b_{q} = e^{i q} b_{q} \bt,\quad \bt b_{q}^+ = e^{-i q} b_{q}^+ \bt\]
and therefore
\be\label{actform_T} \bt  | Q \rangle_{NS} =
e^{{-i \sum_{q\in Q} q} }| Q \rangle_{NS}\,,\qquad
 {}_{NS}\langle Q| \bt =  {}_{NS} \langle Q | e^{{-i \sum_{q\in Q} q} }\,.
\ee

The formulas \rr{vacI} and \rr{bact} allow to find explicit expressions for the eigenvectors
of $H$ as linear combinations of the eigenvectors of $H_0$:
\be\label{QpQ0a}
|Q_+,Q_0\rangle_{NS}=\prod_{q\in Q_+} \left(c_q a_{-q}^+a_{q}^+ + i s_q\right)
   \prod_{q \in  {\bar Q}_+}\left(c_q + i s_q
a_{-q}^+ a_{q}^+\right)\prod^{\longrightarrow}_{q \in Q_0}a_q^+|0\rangle_{0,NS}\,.
\ee
It is convenient to introduce the notations: $\varepsilon(q) =  \varepsilon_q$ if $q\in \ql$ and
$\varepsilon(q) = -\varepsilon_q$ if $q\in \bql$,
\be\label{alphabeta}
\alpha(q) =  \sqrt{\frac{\left(\eps_\pi+\varepsilon(q) \right)\left(\eps_0+\varepsilon(q) \right)}
{4 \varepsilon(q)}}\,,\qquad
\beta(q)= \frac{1}{2\varepsilon(q)}\sqrt { {\left(\eps_\pi-\varepsilon(q) \right)
\left(\varepsilon(q)-\eps_0\right)}{\varepsilon(q)}}\,.
\ee
The formulas \rr{meps} give
\[
\alpha(q)=c_q,\,\beta(q)=s_q\,\,\mbox{if}\, \, \, q\in \ql\,;
\qquad \alpha(q)=s_q, \,\beta(q)=-c_q \,\,\mbox{if}\, \, \,q \in \bql\,,
\]
which allow to present the products over $Q_+$ and $\bar Q_+$ in \rr{QpQ0a} uniformly:
\[
| Q_+,Q_0\rangle_{NS} =i^{\bar m_+}\prod_{q\in {\cal Q}_+} \left(\alpha(q) a_{-q}^+a_{q}^+ + i \beta(q) \right)
   \prod^{\longrightarrow}_{q \in Q_0}a_q^+|0\rangle_{0,NS}=
\] \be\label{QINS}
= i^{m_+}(-1)^{{\bar m}_+}\sum_{\ql'\subset {\cal Q}_+} i^{-m'_+}\prod_{q\in\bql'}
\beta(q) \prod_{q\in \ql'}
 \alpha(q) \,  |\ql', Q_0\rangle_{0,NS},
\ee
where
\[
|\ql', Q_0\rangle_{0,NS}=
\prod_{q\in \ql'}( a_{-q}^+a_{q}^+)\prod^{\longrightarrow}_{q \in Q_0}a_q^+ |0\rangle_{0,NS}.
\]
It means that the eigenvectors of the Hamiltonian
$H$ in given Onsager sector ${\cal V}_{Q_0}$ can be represented as linear combinations of the eigenvectors of the Hamiltonian $H_0$
from the same  Onsager sector.

Similarly  to \rr{QINS} the eigenvectors of the Hamiltonian $H$  from $R$-sector are
\be\label {PIR}
|P_+,P_0\rangle_{R}
= i^{n_+}(-1)^{{\bar n}_+}\sum_{\ppl'\subset \cpl} i^{-n'_+}\prod_{p\in\bpl'}
\beta(p) \prod_{p\in \ppl'}
 \alpha(p)     |\ppl', P_0\rangle_{0,R},
\ee
where
\[
|\ppl', P_0\rangle_{0,R}=\prod_{p\in \ppl'}( a_{-p}^+a_{p}^+)\prod^{\longrightarrow}_{p \in P_0}a_p^+ |0\rangle_{0,R}\,.
\]

\section{Spin operator matrix elements for the eigenstates of Hamiltonian  $H$}

For calculation of  the matrix element
  ${}_{R}\langle {\cp}|\bx_1 |{\cq}\ra_{NS}$ we use the expressions
  (\ref{QINS}) and (\ref{PIR}) for the eigenstates of the Hamiltonian  $H$ and (\ref{mat_fin_0})
  for the matrix element   ${\, }_{0,R}\langle  \po , \ppl'| \bx_1| {\ql', \qo }\rangle_{0,NS}$.
  In order to obtain the factorized formula for the matrix elements   ${}_{R}\langle {\cp}|\bx_1 |{\cq}\ra_{NS}$
we need to make summation over the eigenstates of $H_0$ labeled by the sets $\ql'$ and $\ppl'$. For the summation we will
do some algebraic transformations to fit the summation formula from Appendix B:
\[
  {}_{R}\langle  \po , \ppl| \bx_1| {\ql, \qo }\rangle_{NS}=i^{m_+-n_+}(-1)^{{\bar n}_++{\bar m}_+}
  \sum_{\ql'\subset {\cal Q}_+}  \sum_{\ppl'\subset {\cal P}_+} i^{n'_+-m'_+}
\times
\]\[
\times \prod_{p\in\bpl'}
\beta(p) \prod_{q\in\bql'}\beta(q) \prod_{p\in \ppl'}\alpha(p)\,\,\prod_{q\in \ql'}\alpha(q)\,\cdot
{}_{0,R}\langle   \po , \ppl'| \bx_1| {\ql', \qo }\rangle_{0,NS}=
\]\[
=i^{m_+-n_+ }(-1)^{{\bar n}_+ +{\bar m}_+}  N_{\po,\qo}
  \sum_{\ql'\subset {\cal Q}_+}  \sum_{\ppl'\subset {\cal P}_+}  \delta_{|I'|,\mu} \times
\]\[
\times\prod_{p\in  {\cal P}_+}\beta(p)\prod_{q\in {\cal Q}_+}\alpha(q)
\prod_{p\in\ppl'} \frac{-\alpha(p)}{\beta(p)}\tan\frac{p}{2}
\prod_{q\in \bql'}\frac{\beta(q)}{\alpha(q)}\tan\frac{q}{2}\times
\]\[\times \frac{1}{\prod_{x\in I'}(\cos x+1)^\sigma
\prod_{y\in \bar I'}(\cos y-1)^\tau\prod_{x\in I',y\in \bar I'}(\cos x-\cos y)},\]
where
$I' = \ql'\cup\bpl'$, $\bar I' = \ppl'\cup\bql'$,
\be\label{mumup}
\mu =\frac 12\left (L-m_0-{\bar n}_0 \right), \qquad \bar\mu = \frac 12\left(L-n_0-{\bar m}_0 \right)\,.
\ee
The following relations will be useful in what follows:
\be\label{relmu}
\tau+\mu=\sigma+\bar \mu\,,
\quad |I|=\mu+\frac{m-n}{2}\,,\quad
\quad |\bar I|=\bar \mu+\frac{n-m}{2}\,,
\ee
where definitions of $\sigma$ and $\tau$ are given by \rr{defsigtau}. Using the relations
\be\label{albe2}
\left|\tan\frac{p}{2}\right|=\sqrt{\frac{\varepsilon(p)^2-\eps_0^2}{\eps_\pi^2-\varepsilon(p)^2}}\,,\quad
\frac{\alpha(p)}{\beta(p)}\cdot \left|\tan\frac{p}{2}\right|  =\frac{ \eps_0+\varepsilon(p)}{\eps_\pi-\varepsilon(p)}\,,\quad
 -\frac{\beta(q)}{\alpha(q)}\cdot\left|\tan\frac{q}{2}\right|=\frac{\eps_0-\varepsilon(q)}{\eps_\pi+\varepsilon(q)}
\ee
and
\be\label{cos1cos2}
 \cos x+1 = \frac{\eps_\pi^2-\varepsilon(x)^2}{2\kp},\quad  \cos x-1 = \frac{\eps_0^2-\varepsilon(x)^2}{2\kp},
\quad \cos x-\cos y = \frac{\varepsilon(y)^2-\varepsilon(x)^2}{2\kp},
\ee
  in the expression for the matrix element we get
\[
{}_{R}\langle  \po , \ppl| \bx_1| {\ql, \qo}\rangle_{NS}=
i^{ m_+-n_+}(-1)^{{\bar n}_++  \bar m_++\bar\mu}
 N_{\po,\qo}\prod_{p\in {\cal P}_+}\beta(p)\prod_{q\in{\cal Q}_+}\alpha(q)\ \times
\]\[
\times\sum_{\ppl, \ql} \delta_{|I'|,\mu}
\prod_{p\in\ppl'}\frac{\eps_0+\varepsilon(p)}{\eps_\pi-\varepsilon(p)}
\prod_{q\in \bql'}\frac{ \eps_0-\varepsilon(q)}{\eps_\pi+\varepsilon(q)}\times
\]\[\times
\frac{(2\kp)^{\sigma|I'|+\tau|\bar I'|+|I'||\bar I'|}}
{\prod_{x\in I'}\left(\eps_\pi^2-\varepsilon(x)^2\right)^\sigma
\prod_{y\in \bar I'}\left(\eps_0^2-\varepsilon(y)^2\right)^\tau
\prod_{x\in I',y\in \bar I'}\left(\varepsilon(y)^2-\varepsilon(x)^2\right)},
\]

In the terms of the notations
\[
\gamma_p =- \varepsilon(p),\,\,\mbox{if}\, \, \,p\in \cpl,
\quad \gamma_q = \varepsilon(q),\,\,\mbox{if}\, \, \, q\in \cql,
\]
we have
\[
{}_{R}\langle  \po , \ppl| \bx_1| {\ql, \qo }\rangle_{NS}=
\]\[
 =i^{m_+-n_+ }(-1)^{{\bar n}_++\bar m_++\bar\mu}N_{\po,\qo}
\frac{(2\kp)^{\sigma\mu+\tau\bar\mu+\mu\bar\mu}}
{\prod_{x\in \cpl \cup \cql}\left(\eps_\pi^2-\gamma_x^2\right)^\sigma}
\prod_{p\in \cpl}\beta(p)\prod_{q\in\cql}\alpha(q)\times
\]\[
\times\sum_{I'\subset \cpl \cup \cql} \delta_{|I'|,\mu}
\prod_{y\in \bar I'} \frac{\left(\eps_\pi-\gamma_y\right)^\sigma\left(\eps_0-\gamma_y\right)^{1-\tau}}
{\left(\eps_\pi+\gamma_y\right)^{1-\sigma}\left(\eps_0+\gamma_y\right)^\tau}\cdot
\frac{1}{\prod_{x\in I',y\in \bar I'}\left(\gamma_y^2-\gamma_x^2\right)}.
\]

Now we make summation over $I'$  by means of the following formula from Appendix B:
\[
\sum_{I'\subset \cpl \cup \cql} \delta_{|I'|,\mu}
\prod_{y\in \bar I'} \frac{\left(\eps_\pi-\gamma_y\right)^\sigma\left(\eps_0-\gamma_y\right)^{1-\tau}}
{\left(\eps_\pi+\gamma_y\right)^{1-\sigma}\left(\eps_0+\gamma_y\right)^\tau}\cdot
\frac{1}{\prod_{x\in I',y\in \bar I'}\left(\gamma_y^2-\gamma_x^2\right)}=
\]\[
=\frac{2^{\min(\mu,\bar\mu)}(-1)^{\mu(\mu+1)/2}(-1)^{\sigma(1-\tau)}}{\prod_{x\in
\cpl \cup \cql}\left(\eps_\pi+\gamma_x\right)^{1-\sigma}\left(\eps_0+\gamma_x\right)^\tau\prod_{x<y\in
\cpl \cup \cql}\left(\gamma_x+\gamma_y\right)}\,,
\]
where $\min(\mu,\bar\mu) = (\mu+\bar\mu-|\mu-\bar\mu|)/2=(|I|+|\bar I|-|\sigma-\tau|)/2$. As a result we obtain
\[
{}_R\langle  \po , \ppl| \bx_1| {\ql, \qo }\rangle_{NS}=
i^{m_+-n_+ }(-1)^{{\bar n}_++\bar m_++\bar\mu}N_{\po,\qo}
\frac{(2\kp)^{\sigma\mu+\tau\bar\mu+\mu\bar\mu}}{\prod_{x\in \cpl \cup \cql}\left(\eps_\pi^2-\gamma_x^2\right)^\sigma}\times
\]\[
 \prod_{p\in \cpl}\beta(p)\prod_{q\in\cql}\alpha(q)\,
 \frac{2^{(|I|+|\bar I|-|\sigma-\tau|)/2}(-1)^{\mu(\mu+1)/2}(-1)^{\sigma(1-\tau)}}{\prod_{x\in
 \cpl \cup \cql}\left(\eps_\pi+\gamma_x\right)^{1-\sigma}\left(\eps_0+\gamma_x\right)^\tau
 \prod_{x<y\in  \cpl \cup \cql}\left(\gamma_x+\gamma_y\right)} \]
Using the relations (\ref{alphabeta}),  (\ref{mumup}), \rr{relmu} and
$\gamma_\alpha = \eps_\alpha$ for $\alpha\in I$, $\gamma_\beta = -\eps_\beta$ for $\beta\in \bar I$ we get
\[
{}_R\langle  \po , \ppl| \bx_1| {\ql, \qo }\rangle_{NS}=i^{m_+-n_+ } (2\kp)^{\sigma\mu+\tau\bar\mu+\mu\bar\mu} N_{\po,\qo}M_-
 \times
\]\[2^{-|\sigma-\tau|/2}
\prod_{\alpha\in I}\sqrt{\frac{(\eps_\pi+\eps_\alpha)(\eps_\alpha+\eps_0)}{2\eps_\alpha}}
\prod_{\beta\in \bar I}\sqrt{\frac{(\eps_\pi-\eps_\beta)(\eps_\beta-\eps_0)}{2\eps_\beta}}
\times\]
\[
 \frac{1 }{\prod_{\alpha\in I}\left(\eps_\pi+\eps_\alpha\right)
 \left(\eps_\pi-\eps_\alpha\right)^\sigma\left(\eps_0+\eps_\alpha\right)^\tau
\prod_{\beta\in \bar I}\left(\eps_\pi-\eps_\beta\right)\left(\eps_\pi+\eps_\beta\right)^\sigma\left(-\eps_\beta+\eps_0\right)^\tau}\times\]
\be\label{f_after_sum}
\frac{1}{\prod_{\alpha<\alpha'\in I}\left(\eps_\alpha+\eps_{\alpha'}\right)\prod_{\beta<\beta'\in \bar I}\left(-\eps_\beta - \eps_{\beta'}\right)
\prod_{\alpha\in I}\prod_{\beta\in \bar I}\left(\eps_\alpha-\eps_\beta\right)}\,,
\ee
where
\[
M_-=(-1)^{\bar m_++\bar\mu+\mu(\mu+1)/2+\sigma(1-\tau)}\,.
\]
Let us substitute here the coefficient $N_{\po,\qo}$ which follows from \rr{idenA1} of Appendix A:
\[
N_{\po,\qo}=\delta^{\,(\text{mod}\, 2)}_{m_0-n_0, 0  }\,i^{n_+ - m_+}2^{-(n-m )^2/4} \left(\frac{2}{L}\right)^{(n+m)/2} N_-\times
\]\[
 \prod_{p\in \po}e^{ip(m-n)/2}\prod_{q\in \qo}e^{-iq(m-n-2)/2}A_{P,Q} \times
\]\be\label{npqH}
(-1)^{n_+}\prod_{\beta\in {\bar I}}\frac{1}{\tan(\beta/2)}
\prod_{\beta\in {\bar I}}(\cos \beta-1)^{\tau}
\prod_{\alpha\in I}(\cos \alpha+1)^{\sigma}\prod_{\alpha\in I}
\prod_{\beta\in {\bar I}}
(\cos \alpha - \cos \beta)\,,\ee
where
\[ A_{P,Q}=\frac{\prod_{p<p'\in P} (\ep-\epp)
\prod_{q<q'\in Q} (\eq-\eqp)}{
\prod_{q\in Q}\prod_{p\in P}
( \eq-\ep)}\,,
\]\[
N_-=(-1)^{n(n-1)/2} (-1)^{(m-n)(m-n-2)/8} (-1)^{(L-m_0-\bar n_0)(n-m)/4}\,.
\]
Then using \rr{cos1cos2} and $2^{-|\sigma-\tau|/2}=2^{-1/4} (\eps_0+\eps_\pi)^{(1-2\sigma)(1-2\tau)/4}$,
which follows from $\sigma,\tau\in\{0,1\}$ and  $\eps_0=1-\kp$, $\eps_\pi=1+\kp$, we get
\[
{}_R\langle  \po , \ppl| \bx_1| {\ql, \qo}\rangle_{NS}=
 \delta^{\,(\text{mod}\, 2)}_{m-n, 0  }
\left(\frac{2}{L}\right)^{(m+n)/2}
   (-1)^{n(n-1)/2}(-1)^{(n-m)/2}  \kp^{(n-m )^2/4}
\times\]
\[ \times \prod_{p\in P}e^{ip(m-n)/2}\prod_{q\in Q}e^{-iq(m-n-2)/2}A_{P,Q}\, \times \]
\[
\times 2^{-1/4} (\eps_0+\eps_\pi)^{(1-2\sigma)(1-2\tau)/4}
\prod_{\alpha\in I}\sqrt{\frac{(\eps_\pi+\eps_\alpha)(\eps_\alpha+\eps_0)}{2\eps_\alpha}}
\prod_{\beta\in \bar I}\sqrt{\frac{1}{2\eps_\beta(\eps_\pi+\eps_\beta)(\eps_\beta+\eps_0)}}
\,\times \]
\[
\times  \frac{\prod_{\beta\in \bar I}(\eps_0+ \eps_\beta)^{\tau}(\eps_\pi+ \eps_\beta)^{1-\sigma} }
{\prod_{\alpha\in I}\left(\eps_\pi+\eps_\alpha\right)^{1-\sigma}\left(\eps_0+\eps_\alpha\right)^\tau}
 \,\frac{\prod_{\alpha\in I,\beta\in \bar I}\left(\eps_\alpha+\eps_\beta\right)}
{\prod_{\alpha<\alpha'\in I}\left(\eps_\alpha+\eps_{\alpha'}\right)
\prod_{\beta<\beta'\in \bar I}\left(\eps_\beta + \eps_{\beta'}\right)}\,,\]
where we also extended the products over $P_0$ and $Q_0$ of the exponents of momenta to $P$ and $Q$, respectively,
since it means addition of pairs of opposite momenta.
Regrouping factors with respect to the sets
$U = Q + \bar P = Q_+ + Q_- + Q_0 + \bar P_+ + \bar P_- + \bar P_0$ and $\bar U = P + \bar Q$
we obtain
\[
{}_R\langle  \po , \ppl| \bx_1| {\ql, \qo}\rangle_{NS}=
 \delta^{\,(\text{mod}\, 2)}_{m-n, 0  }
\left(\frac{2}{L}\right)^{(m+n)/2}
   (-1)^{n(n-1)/2+(n-m)/2}  \kp^{(n-m )^2/4}A_{P,Q}\times
\]\be\label{finU}
\times  \prod_{p\in P}e^{ip(m-n)/2}\prod_{q\in Q}e^{-iq(m-n-2)/2}\,
(\eps_0\eps_\pi)^{1/8}\frac{\prod_{\alpha\in U,\beta\in \bar U}(\eps_\alpha+\eps_\beta)^{1/4}}
{\prod_{\alpha,\alpha'\in U}(\eps_\alpha+\eps_{\alpha'})^{1/8}
\prod_{\beta, \beta'\in \bar U}(\eps_\beta+\eps_{\beta'})^{1/8}}\,.
\ee
To prove this formula we have to compare the exponents of $(\eps_\alpha+\eps_\beta)$
for all  $\alpha$ and $\beta$ in the former and latter expressions and to use definitions \rr{defsigtau} of $\sigma$ and $\tau$.
Now we rewrite the products over the sets $U$ and $\bar U$
as products over the sets $P$, $Q$, $\bar P$, $\bar Q$. Finally we exclude the products over the sets $\bar P$, $\bar Q$
supplementing them to be the products over all the sets of momenta in the first Brillouin zone  ${\cal \cq}=\cq\cup\bar \cq$ for $NS$-sector
and ${\cal \cp}=\cp\cup\bar\cp$ for $R$-sector. It gives the factorized formula for the matrix element of spin operator \cite{Iorgov1}:
\[
{}_{R}\langle  P| \bx_1| {Q}\rangle_{NS}=
{}_{R}\langle {p_1}, \,{p_2}, \ldots , {p_n}|\bx_1 |{q_1},\,
 {q_2}, \ldots, \,{q_m}\ra_{NS} =
 \]\[=\delta^{(\text{mod}\, 2)}_{m-n,0}\,
  i^{-(n+m)/2}(-1)^{n(n-1)/2} \prod_{p\in P}e^{-i p/2}\prod_{q\in Q}e^{i q/2}
\times
\]\[
\times
 \left(\frac{2}{L}\right)^{(m+n)/2}\kp^{(m-n)^2/4} \:\sqrt{\xi\; \xi_T} \;\prod_{{q\in Q}}
 \frac{ e^{\eta(q)/2}}{ \sqrt{2 \varepsilon_q}}\;\:\prod_{{p\in P}}\:
\frac{ e^{-\eta(p)/2}}{ \sqrt{2 \varepsilon_p}}\;\;
\times
\]
\be\label{mat_fin}
\times\prod_{q<q'\in Q}\left(\frac{2\sin \frac{q-q'}{2}}{\varepsilon_q+
\varepsilon_{q'}}\right)
\prod_{p<p' \in P}\left(\frac{2\sin \frac{p-p'}{2}}{\varepsilon_p+\varepsilon_{p'}}\right)
\prod_{{q \in Q}} \; \prod_{p \in P} \:
\left(\frac{\varepsilon_q+\varepsilon_p}{2\sin \frac{q-p}{2}}\right)\,,
\ee
where $\varepsilon_\alpha$ is given by \rr{eeq} and
\[
\xi\:=\: \left( 1-\kp^2\right)^{\frac{1}{4}}\,,\quad
\xi_T\:=\:\frac{\prod_{q\in{\cal \cq}}\; \prod_{p\in{\cal \cp}}
(\varepsilon_q+\varepsilon_p)^{\frac{1}{2}}}{\prod_{q,q'\in{\cal \cq}}\;(
\varepsilon_q+ \varepsilon_{q'})^{\frac{1}{4}}\;\prod_{p,p'\in{\cal \cp}}\; (\varepsilon_p+
\varepsilon_{p'})^{\frac{1}{4}}}\,,\quad
e^{\eta(\alpha)}=\frac{ \prod_{q' \in {\cal \cq}}\left(
\varepsilon_\alpha+\varepsilon_{q'}\right)}{\prod_{p'\in {\cal \cp}} \left(\varepsilon_\alpha+\varepsilon_{p'}\right)}\,.
\]
Matrix elements with even (resp. odd) $n$ and $m$ correspond to the Hamiltonian \rr{hamP} for periodic boundary condition
(resp. \rr{hamA} for antiperiodic boundary condition).

Let us comment on the modification of formula \rr{mat_fin} in the case of the spin $\bx_k$ matrix elements.
{}From (\ref{actform_T}) and similar formula for R-sector and also from \rr{tran_act} we obtain
\[ {}_R\langle  P| \bx_k| Q\rangle_{NS} = {}_R\langle P| T^{k-1}\bx_1
T^{-k+1} | Q \rangle_{NS} =  e^{ i(k-1)\left(\sum_{q\in Q} q - \sum_{p\in P} p\right)} {}_R\langle
P|\bx_1 | Q \rangle_{NS}.\]

\section{Conclusions}

In the  paper \cite{gipstZN}  using Baxter's extension of the
Onsager algebra  the  factorized expressions for the spin operator matrix elements  between the
eigenstates of
Hamiltonians of the finite length superintegrable $Z_N$-symmetric chiral Potts quantum chain (SCPS) were found
up to unknown scalar factors  for any pair of the Onsager sectors.
In this paper we have derived the exact expression \rr{npqA} for these factors
for  quantum Ising chain in a transverse field ($N=2$ SCPC-model).
This derivation uses some modification of the standard fermion technique \cite{Lieb}.
On the first  stage  we diagonalized the Hamiltonian $H_0$ (\ref{ham01}) by means of the dual
Jordan--Wigner fermion operators and calculated the spin matrix elements between  eigenstates of $H_0$,
then after the Bogoliubov transformation we obtained
the factorized formula for the spin matrix elements between iegenvectors of $H$ using the summation formula (Appendix B).
It is natural to expect that the fermion technique used in this paper
can be applied to calculation of the spin matrix elements in more general free fermion models, for example,
in the $N=2$ Baxter--Bazhanov--Stroganov model \cite{buiorgov,Iorgov2}.
We will address this problem in our forthcoming paper.

\medskip

\noindent {\bf Acknowledgements.} We thank  G.~von~Gehlen, O.~Lisovyy, S.~Pakuliak for useful discussions.
This work was supported by the Program of Fundamental Research of the Physics and Astronomy Division of the NAS of Ukraine, the Ukrainian FRSF
grants $\Phi$28.2/083 and $\Phi$29.1/028, by French-Ukrainian program “Dnipro” M17-2009 and the joint project PICS
of CNRS and NAS of Ukraine. N.I. is thankful to Max-Planck-Institut f\"ur Mathematik for kind hospitality.

\section*{Appendix A}

In this Appendix we prove the following trigonometric identity for the momenta of quantum Ising chain.
We will use the definitions from Sect.~4.2 for different sets of momenta.
Supposing that $m=|Q|$, $n=|P|$, $m_0=|Q_0|$ and $n_0=|P_0|$ have equal parity, we claim that
\[ A_{P,Q}:=\frac{\prod_{p<p'\in P} (\ep-\epp)
\prod_{q<q'\in Q} (\eq-\eqp)}{
\prod_{q\in Q}\prod_{p\in P}
( \eq-\ep)} =
\]\[
= \delta^{(\text{mod}\, 2)}_{m_0-n_0,0} \left(\frac{2}{L}\right)^{-(n+m)/2}  i^{(n_0-m_0)/2}i^{(m-n)/2}2^{(n-m)^2/4}
     \prod_{p\in P_0}e^{ ip(n-m)/2 }\prod_{q\in Q_0}e^{ iq(m-n-2)/2 }\times
\]
\be\label{idenA1}\times N_-  N_{P_0,Q_0}
\frac{(-1)^{n_+}
\prod_{\beta\in {\bar I}}\tan{\frac{\beta}{2}}}
{\prod_{\alpha\in I}(\cos\alpha+1)^\sigma\prod_{\beta\in
{\bar I}}(\cos\beta-1)^\tau\prod_{\alpha\in I,\beta\in {\bar I}}(\cos\alpha-\cos\beta)}\,,
\ee
with $I =\ql\cup\bpl$, ${\bar I} = \ppl\cup\bql$,
\[
\delta^{\,(\text{mod}\, 2)}_{m_0-n_0, 0  } = \frac 12\left((-1)^{n_0-m_0}+1\right)\,,\quad
N_-=(-1)^{n(n-1)/2+(m-n)(m-n-2)/8+(L-m_0-\bar n_0)(n-m)/4}\,,
\]
and the coefficient $N_{\po,\qo}$ not depending on $P_+$ and $Q_+$ (or equivalently not depending on $I$).
In the main text we use this formula for different choices of $P_+$ and $Q_+$, namely,
as labels of eigenvectors of $H_0$  or as labels of eigenvectors of $H$ with $P_0$ and $Q_0$ being fixed.

In order to prove (\ref{idenA1}) it is sufficient
to show that addition of a pair of opposite momenta to $P$ or to $Q$ in the left-hand side of (\ref{idenA1})
does not change the coefficient $N_{\po,\qo}$ in the right-hand side of (\ref{idenA1}).
Let us add, for example, a pair of momenta $\{-p,p\}$ to $P$ and calculate the ratio of left-hand sides of
(\ref{idenA1}):
\be\label{Aratio}
\frac{A_{P\cup \{-p,p\},Q}}{A_{P,Q}}=\frac{\prod_{p'\in \po}B(p,p')}{\prod_{q'\in \qo}B(p,q')}\cdot
\frac{\prod_{p'\in P_+}C(p,p')}{\prod_{q'\in Q_+}C(p,q')}\cdot(e^{-ip}-e^{ip})\,,
\ee
where
\[B(\alpha,\beta)=(e^{i\alpha}-e^{i\beta})(e^{-i\alpha}-e^{i\beta})\,,\quad
C(\alpha,\beta)=B(\alpha,\beta)B(\alpha,-\beta)\,.
\]
 Note that in (\ref{idenA1}) we  used the following ordering of the momenta
  $\{{q_1},\, {q_2}, \ldots, \,{q_m}\}$:
  from the beginning the pairs $(-q, q)$, for $q\in Q_+$, are going on and then $q\in Q_0$ in a order fixed in all the paper.
Analogously we define the ordering of the momenta in the set $P=\{{p_1},\, {p_2}, \ldots, \,{p_n}\}$.
It order to present the factors of \rr{Aratio} in convenient form we introduce the following functions:
\[
\chi(\alpha,\beta)=\left(1-e^{i(\alpha+\beta)}\right)\left(1-e^{i(\alpha-\beta)}\right)\,, \quad
 \varphi_+(\alpha)=1+e^{i\alpha}\,, \quad \varphi_-(\alpha)=
1-e^{i\alpha}\,.
\]
For the products of the functions  $B$  we use the first two of following identities:
\be\label{id1ch3}
  \prod_{p'\in \po}B(p,p') =
L(-1)^{n_0}e^{-i p n_0}\,\prod_{p'\in \po}\epp\, \frac{\varphi_-(p)^{P_{0,0}-{\bar P}_{0,0}-1}
\varphi_+(p)^ {P_{0,\pi}-{\bar P}_{0,\pi}-1}}
{\prod_{p'\in {\cal P}_+}^{p'\neq p}\chi(p,p')},
\ee
\be\label{id2ch3}
  \prod_{q'\in \qo}B(p,q')  =
2(-1)^{m_0}e^{-i p m_0}\,\prod_{q'\in \qo}\eqp\,
 \frac{\varphi_-(p)^{Q_{0,0}-{\bar Q}_{0,0}} \varphi_+(p)^{Q_{0,\pi}-{\bar Q}_{0,\pi}}} {\prod_{q'\in {\cal Q}_+}\chi(p,q')},
\ee
\be\label{id3ch3}
   \prod_{q'\in \qo}B(q,q') =
L(-1)^{m_0}e^{-i q m_0} \,\prod_{q'\in \qo}\eqp\,
 \frac{\,\varphi_-(q)^{Q_{0,0}-{\bar Q}_{0,0}-1}\varphi_+(q)^
{Q_{0,\pi}-{\bar Q}_{0,\pi}-1}} {\prod_{q'\in {\cal Q}_+}^{q'\neq q}\chi(q,q')},
\ee
\be\label{id4ch3}
  \prod_{p'\in \po}B(q,p')
  = 2(-1)^{n_0}e^{-i q n_0}\,\prod_{p'\in \po}\epp\,
\frac{\,\varphi_-(q)^{P_{0,0}-{\bar P}_{0,0}}\varphi_+(q)^{P_{0,\pi}-{\bar P}_{0,\pi}}}
{\prod_{p'\in {\cal P}_+}\chi(q,p')}.
\ee
The functions $P_{0,0}$, $\bar P_{0,0} $, $P_{0,\pi}$, $\bar P_{0,\pi}$ are defined by \rr{defP0Ppi}.
The functions $Q_{0,0}$, $\bar Q_{0,0}$, $Q_{0,\pi}$, $\bar Q_{0,\pi} $ are defined similarly.
Note that always we have $Q_{0,0}=\bar Q_{0,0}=0$ but we included them for uniformity.
The identities (\ref{id1ch3})--(\ref{id4ch3}) follow  from the elementary trigonometric identities:
\[\prod_{k=1}^{L-1}\left(1-\om^k\right) = L,\quad
 \prod_{k=1}^{L}\left(1-\om^{k+1/2}\right) =   2, \quad \om = e^{2\pi i/L}.
\]
We also have
\[
C(\alpha,\beta)=e^{-2i\alpha} \chi(\alpha,\beta)^2\,,\quad
e^{-ip}-e^{ip}=e^{-ip}\varphi_+(p)\varphi_-(p)\,.
\]
Since $m_0$ and $n_0$ has the same parity, for the ratio \rr{Aratio} we have
\[
\frac{A_{P\cup \{-p,p\},Q}}{A_{P,Q}}=
\frac{L}{2}  e^{i p (m-n-1)}\,\prod_{p'\in \po}e^{ip'}\,
\prod_{q'\in \qo}e^{-iq'}\cdot
 \frac{\varphi_-(p)}{\varphi_+(p)}\,\frac{\varphi_+(p)^{2\sigma}}{\varphi_-(p)^{2\tau}}\,
\frac{\prod_{\beta \in \bar I}\chi(p,\beta)}{\prod_{\alpha\in I}^{\alpha \neq p}\chi(p,\alpha)}\,,
\]
where we used \rr{defsigtau} and the definition of the sets $I$ and $\bar I$. Taking into account the identities
\[
e^{-i \alpha}\chi(\alpha,\beta)=
2(\cos\alpha-\cos\beta),\quad
e^{-i \alpha}\varphi^2_+(\alpha)=
2(\cos\alpha+1),\]
\[
e^{-i \beta}\varphi^2_-(\beta)=
2(\cos\beta-1),\quad\frac{\varphi_-(\alpha)}{\varphi_+(\alpha)}= -i\tan(\alpha/2)
\]
and the relation $|\bar I|-|I|+\sigma - \tau=n-m$ following from \rr{relmu} we get
\[
\frac{A_{P\cup \{-p,p\},Q}}{A_{P,Q}}=-i
\frac{L}{2}\, 2^{n-m+1}\,\prod_{p'\in \po}e^{ip'}\,
\prod_{q'\in \qo}e^{-iq'}\,  \times
\]\be\label{finrat}
\tan\frac{p}{2}\,\frac{(\cos p+1)^{\sigma}}{(\cos p-1)^{\tau}}\,
\frac{\prod_{\beta \in \bar I}(\cos p- \cos \beta)}
{\prod_{\alpha\in I}^{\alpha \neq p}(\cos p- \cos \alpha)}\,.
\ee
The corresponding ratio of right-hand sides of \rr{idenA1} gives the same result \rr{finrat}.
For the ratio $A_{P,Q\cup \{-q,q\}}/{A_{P,Q}}$ the calculation goes in the same way with the use
\rr{id3ch3} and \rr{id4ch3} for the products of $B$. It proves the identity \rr{idenA1}.

\section*{Appendix B}
The following summation formula over the subsets $I\in {\cal R}$, ${\cal R}=\cpl \cup \cql$:
\[
\sum_{I\subset {\cal R}} \delta_{|I|,\mu}
\prod_{y\in \bar I} \frac{\left(\eps_\pi-\gamma_y\right)^\sigma\left(\eps_0-\gamma_y\right)^{1-\tau}}
{\left(\eps_\pi+\gamma_y\right)^{1-\sigma}\left(\eps_0+\gamma_y\right)^\tau}\cdot
\frac{1}{\prod_{x\in I,y\in \bar I}\left(\gamma_y^2-\gamma_x^2\right)}=
\]\be\label{sumform}
=\frac{2^{\min(\mu,\bar\mu)}(-1)^{\mu(\mu+1)/2}(-1)^{\sigma(1-\tau)}}{\prod_{x\in
{\cal R}}\left(\eps_\pi+\gamma_x\right)^{1-\sigma}\left(\eps_0+\gamma_x\right)^\tau\prod_{x<y\in
{\cal R}}\left(\gamma_x+\gamma_y\right)}\,,
\ee
where $\min(\mu,\bar\mu) = (\mu+\bar\mu-|\mu-\bar\mu|)/2$, is valid.
In fact the formula contains four subcases depending on the choice of $\sigma,\tau\in\{0,1\}$.
In this formula $\mu$ is defined from the relations $\mu+\tau=\bar\mu+\sigma$, $\mu+\bar\mu=|{\cal R}|=r$.
It is easy to prove this formula reducing it
to the summation formulas from Appendix A of \cite{gipstZN} for arbitrary variables $u$, $v$, $z_a$, $a=1,\ldots,r$ and even and odd $r$, respectively:
\be\label{sum_even}
\sum_{I\subset \cal R} \frac{\delta_{|I|,r/2}}{\prod_{a\in I}(z_a+u)\prod_{b\in \bar I} (z_b+v) \prod_{a\in I, b\in \bar I} (z_a^2-z_b^2)}=
\frac{(-1)^{r(r-2)/8} (u-v)^{r/2}}{\prod_{c} (z_c+u)(z_c+v) \prod_{c<s} (z_s+z_c)}\,,
\ee\be\label{sum_odd}
\sum_{I\subset \cal R} \frac{\delta_{|I|,(r+1)/2}}{\prod_{a\in I}(z_a+u)\prod_{b\in  I} (z_b+v) \prod_{a\in I, b\in \bar I} (z_a^2-z_b^2)}=
\frac{(-1)^{(r+1)(r-1)/8} (u+v)^{(r-1)/2}}{\prod_{c} (z_c+u)(z_c+v) \prod_{c<s} (z_s+z_c)}\,.
\ee
We may consider four cases of $\sigma,\tau\in\{0,1\}$ one by one.

For example let us prove \rr{sumform} for the case  $\sigma=1$, $\tau=0$.
In this case $\mu=(r+1)/2$, $\bar\mu=(r-1)/2$. The left-hand side of \rr{sumform} is
\[
\sum_{I\subset {\cal R}}
\frac{\delta_{|I|,(r+1)/2}\prod_{y\in \bar I} \left(\eps_\pi-\gamma_y\right)\left(\eps_0-\gamma_y\right)}
{\prod_{x\in I,y\in \bar I}\left(\gamma_y^2-\gamma_x^2\right)}=
\prod_{y\in {\cal R}} \left(\gamma_y-\eps_\pi\right)\left(\gamma_y-\eps_0\right)\times
\]\[
\times \sum_{I\subset {\cal R}} \frac{\delta_{|I|,(r+1)/2}\,(-1)^{\mu \bar\mu}}
{\prod_{y\in I} \left(\gamma_y-\eps_\pi\right)\left(\gamma_y-\eps_0\right)
\prod_{x\in I,y\in \bar I}\left(\gamma_x^2-\gamma_y^2\right)}=
\frac{2^{(r-1)/2}(-1)^{\mu(\mu-1)/2+\mu-1}}{\prod_{x<y\in
{\cal R}}\left(\gamma_x+\gamma_y\right)}\,,
\]
where we used \rr{sum_odd} for  $u=-\eps_0$, $v=-\eps_\pi$, $\{z_a\}=\{\gamma_a\}$ and $u+v=-2$.
Thus we proved \rr{sumform} for the case  $\sigma=1$, $\tau=0$. The other three cases of $\sigma$ and $\tau$
can be considered similarly.

\end{document}